\documentstyle[12pt]{article}
\begin{document}
\baselineskip 24pt
\hfuzz=1pt
\setlength{\textheight}{8.5in}
\setlength{\topmargin}{0in}
\begin{center}
\Large {\bf  A Bell inequality which can be used to test locality
more simply than 
Clauser-Horne inequality
and which is violated by a larger magnitude of violation than
Clauser-Horne-Shimony-Holt inequality
} \\  \vspace{.75in}
\large {M. Ardehali}\footnote[1]
{email address:ardehali@mel.cl.nec.co.jp}
\\ \vspace{.3in}
Research Laboratories,
NEC Corporation,\\
Sagamihara,
Kanagawa 229
Japan

\end{center}
\vspace{.20in}

\begin{abstract}
A correlation inequality is derived from local realism and a
supplementary assumption. 
Unlike Clauser-Horne (CH) inequality [or
Clauser-Horne-Shimony-Holt (CHSH) inequality]
which is violated by quantum 
mechanics by a factor of 
$\sqrt 2$, this inequality is violated by a factor of $1.5$.
Thus the magnitude of violation of this inequality 
is approximately $20.7\%$ larger than the magnitude of violation of
previous inequalities.
Moreover, unlike 
CH (or CHSH) inequality
which requires the measurement
of five detection probabilities, the present inequality requires the
measurement of only two detection probabilities.
This inequality can therefore be used to test locality more simply 
than CH or CHSH inequality.
\end {abstract}
\pagebreak

Local realism is a philosophical view which holds that external
reality exists and has local properties.
Quantum mechanics vehemently denies that
such a world view has any meaning for physical systems
because local realism
assigns simultaneous values to non-commuting observables.
In $1965$ Bell \cite{1} showed that the assumption  of local realism, as
postulated by Einstein, Podolsky, and Rosen (EPR) \cite{2},
leads to some
constraints on the statistics of two spatially separated particles.
These constraints, which are collectively known as Bell inequalities,
are sometimes grossly violated by quantum mechanics. The violation of
Bell inequalities therefore indicate that local realism is not only
philosophically but also numerically
incompatible with quantum mechanics.
Bell's theorem is of paramount importance for undersanding the
foundations of quantum mechanics because it rigorously formulates
EPR's assumption of locality and shows that
all realistic interpretations of quantum mechanics must be nonlocal.

Bell's original argument, however,
can not be experimentally tested
because it relies on perfect correlation of the spin of the two
particles \cite {3}. Faced with this problem,
Clauser-Horne-Shimony-Holt (CHSH) \cite{4},
Freedman-Clauser (FC) \cite{5}, and Clauser-Horne (CH) \cite{6}
derived correlation
inequalities for
systems which do not achieve $100\%$ correlation,
but which do achieve a necessary minimum correlation. 
Quantum mechanics
violates these inequalities
by as much as $\sqrt 2$.
An experiment based on CHSH, or FC,
or CH inequality utilizes one-channel
polarizers in which the dichotomic choice is between the detection of
the photon and its lack of detection. A better experiment is
one in which
a truly binary choice
is made between the ordinary and the extraordinary rays [7-10].
In this letter, we derive a correlation inequality
for two-channel polarizer systems and we show that 
quantum mechanics violates this inequality 
by a factor of $1.5$. Thus the magnitude of
violation of the inequality derived in this paper
is approximately $20.7\%$ larger than
the magnitude of violation of previous
inequalities of [4-10]. Moreover, we show that
unlike CH (or CHSH) inequality
which requires the measurements
of five detection probabilities,
the present inequality
requires the
measurement of only two detection probabilities.
This inequality can therefore
be used to test locality more simply
than CH (or CHSH) inequality.
This result can be particularly
important for the experimental test of local realism.

We start by considering Bohm's \cite{11} version of EPR experiment
in which an unstable source emits pairs of photons in a singlet
state $\mid\!\!\Phi\rangle$. The source is viewed by two
apparatuses.
The first (second) apparatus consists of a polarizer
$P_1 \left(P_2 \right)$
set at angle $\mbox{\boldmath $a$} \left(
\mbox{\boldmath $b$} \right)$,
and two detectors
$D_{1}^{\,\pm} \left (D_{2}^{\,\pm} \right)$
put along the ordinary and the extraordinary beams.
During a period of time $T$, the source emits, say, $N$ pairs of
photons. Let $N^{\,\pm\,\pm}\left(\mbox{\boldmath $a,b$}\right)$
be the number of simultaneous counts
from detectors $D_{1}^{\pm}$ and $D_{2}^{\pm}$,
$N^{\,\pm}\left(\mbox{\boldmath $a$}\right)$
the number of counts from detectors
$D_1^\pm$, and
$N^{\,\pm}\left(\mbox{\boldmath $b$}\right)$
the number of counts from detectors
$D_2^\pm$.
If the time $T$ is sufficiently
long, then the ensemble probabilities
$p^{\;\pm\;\pm}\left(\mbox{\boldmath $a,b$}\right)$ are defined as
\begin{eqnarray}{\nonumber}
p^{\;\pm\;\pm} \left(\mbox{\boldmath $a,b$} \right)&=&
\frac{N^{\;\pm\;\pm} \left(\mbox{\boldmath $a,b$} \right)}{N}, \\ 
\nonumber
p^{\;\pm}(\mbox{\boldmath $a$})&=&
\frac{N^{\;\pm}(\mbox{\boldmath $a$})}{N}, \\ 
p^{\;\pm}(\mbox{\boldmath $b$})&=&
\frac{N^{\;\pm}(\mbox{\boldmath $b$})}{N}.
\end{eqnarray}
\noindent We consider a particular pair of photons and specify its
state with a parameter $\lambda$. Following Bell, we do not 
impose any restriction on the complexity of $\lambda$. 
``It is
a matter of indifference
in the following whether $\lambda$ denotes a single variable or
a set, or even a set of functions, and whether the variables are 
discrete or continuous \cite{1}.''

The ensemble probabilities
in Eq. $(1)$ are defined as
\begin{eqnarray} {\nonumber}
p^{\:\pm\:\pm}(\mbox{\boldmath $a,b$}) &=&
\int p\,(\lambda)\, p^{\;\pm}(\mbox{\boldmath 
$a$} \mid \lambda) \, p^{\;\pm}(\mbox{\boldmath $b$}
\mid \lambda,\mbox{\boldmath $a$}), \\ \nonumber
p^{\:\pm}(\mbox{\boldmath $a$}) &=&
\int p \, (\lambda) \, p^{\;\pm}(\mbox{\boldmath 
$a$} \mid \lambda), \\ 
p^{\:\pm}(\mbox{\boldmath $b$}) &=&
\int p \, (\lambda)\, p^{\;\pm}(\mbox{\boldmath 
$b$} \mid \lambda).
\end{eqnarray}
Equations (2) may be stated in physical terms: The ensemble
probability for detection of photons by
detectors $D^{\;\pm}_{\; 1}$ and $D^{\;\pm}_{\;2}$
[that is $p^{\;\pm\;\pm}(\mbox{\boldmath $a,b$})$]
is equal to the sum or integral of the probability
that the emission is
in the state $\lambda$ [that is $p(\lambda)$], times the conditional
probability that if the emission is in the state $\lambda$,
then a count is triggered by the first detector $D^{\;\pm}_{1}$
[that is $p^{\;\pm}(\mbox{\boldmath $a$}
\mid \lambda)$],
times the conditional probability that 
if the emission is in the state 
$\lambda$ and if the first polarizer is set along axis $\boldmath a$,
then a count is triggered from the second detector $D^{\;\pm}_{2}$
[that is $p^{\;\pm}(\mbox{\boldmath $b$}
\mid \lambda,\mbox{\boldmath $a$})$].
Similarly the ensemble probability for detection of photons by
detector $D^{\;\pm}_{\;1} \left(D^{\;\pm}_{\;2} \right )$
{\large [} that is $p^{\;\pm}(\mbox{\boldmath $a$}) \left
[p^{\;\pm}(\mbox{\boldmath $b$}) \right]$ {\large ]}
is equal to the sum or integral of the probability that the photon
is in the state $\lambda$ [that is $p(\lambda)$], times the
conditional probability that if the
photon is in the state $\lambda$,
then a count is triggered by
detector $D^{\;\pm}_{1} \left(D^{\;\pm}_{2} \right )$
{\large[} that is $p^{\;\pm}(\mbox{\boldmath $a$}
\mid \lambda) \left [p^{\;\pm}(\mbox{\boldmath $b$} \mid 
\lambda ) \right ]$ {\large]}.
Note that Eqs. $(1)$ and $(2)$ are quite general and follow
from the standard rules of probability theory.
No assumption has yet been made that is not satisfied 
by quantum mechanics.

Hereafter, we
focus our attention only on those theories that satisfy
EPR criterion of locality: `` Since at the time of measurement the
two systems no longer interact, no real change can take place in the
second system in consequence of anything that may be done to first
system
\cite {2}''. EPR's criterion of locality can be translated into
the following mathematical equation:
\begin{equation}
p^{\;\pm}(\mbox{\boldmath $b$} \mid \lambda,
\mbox{\boldmath $a$})=
p^{\;\pm}(\mbox{\boldmath $b$} \mid \lambda).
\end{equation}
Equation $(3)$ is the hall mark of local realism.
\footnote [2] {
It is worth noting that there is a
difference between Eq. (3) and CH's
criterion of locality. CH write their assumption of
locality as
\begin{eqnarray*}
p^+ \left(\mbox{\boldmath $a, b$} , \lambda \right)=
p^+ \left(\mbox{\boldmath $a$} , \lambda \right)
p^+ \left(\mbox{\boldmath $b$}, \lambda \right).
\end{eqnarray*}
Apparently by $p^+ \left(\mbox{\boldmath $a, b$} , \lambda \right)$,
they mean the conditional probability that if the
emission is in state $\lambda$,
then simultaneous counts are triggered by detectors
$D^+_1$ and $D^+_2$.  However, what they call
$p^+ \left(\mbox{\boldmath $a, b$} , \lambda \right)$
in probability theory is usually
called $p^+ \left(\mbox{\boldmath $a, b$} \mid\lambda \right)$ [note
that $p (x, y, z)$ is the joint
probability of $x, y$ and $z$, whereas $p (x, y \mid z)$
is the conditional probability that
if $z$ then $x$ and $y$]. Similarly by
$p^+ \left(\mbox{\boldmath $a$} , \lambda \right) {\large [}
p^+ \left(\mbox{\boldmath $b$}, \mid \lambda \right){\large]}$,
CH mean the conditional
probability that if the emission is in state $\lambda$,
then a count is triggered from the
detector $D^+_1 \left(D^+_2 \right)$.
Again what they call
$p^+ \left(\mbox{\boldmath $a$} , \lambda \right) {\large [}
p^+ \left(\mbox{\boldmath $b$}, \lambda \right){\large]}$
in probability
theory is usually written as
$p^+ \left(\mbox{\boldmath $a$} \mid \lambda \right) {\large [}
p^+ \left(\mbox{\boldmath $b$} \mid \lambda \right){\large]}$
(again note that $p(x, z)$ is the
joint probability of $x$ and $z$, whereas $p (x \mid z)$
is the conditional probability
that if $z$ then $x$).
Thus according to standard notation of probability theory,
CH criterion of locality may be written as
\begin{eqnarray*}
p^+ \left(\mbox{\boldmath $a, b$} \mid \lambda \right)=
p^+ \left(\mbox{\boldmath $a$} \mid \lambda \right)
p^+ \left(\mbox{\boldmath $b$} \mid \lambda \right).
\end{eqnarray*}
Now according to Bayes' theorem,
\begin{eqnarray*}
p^+ \left(\mbox{\boldmath $a, b$} \mid \lambda \right)=
p^+ \left(\mbox{\boldmath $a$} \mid \lambda \right)
p^+ \left(\mbox{\boldmath $b$} \mid \lambda,
\mbox{\boldmath $a$}, \right).
\end{eqnarray*}
Substituting the above equation in CH's criterion of locality,
we obtain
\begin{eqnarray*}
p^+ \left(\mbox{\boldmath $ b$}
\mid \lambda, \mbox{\boldmath $a$} \right)=
p^+ \left(\mbox{\boldmath $b$} \mid \lambda\right),
\end{eqnarray*}
which for the ordinary equation is the same as Eq. (3).}
It is the most general form of locality that accounts
for correlations subject only to the requirement that a count
triggered by the second detector does not depend on
the orientation of the first polarizer. The assumption
of locality, i.e., Eq. $(3)$, is
quite natural since the two photons are spatially separated so that
the orientation of the first polarizer should not influence the
measurement carried out on the second photon.

In the following we show that 
equation $(3)$ leads to validity of an equality
that is sometimes grossly violated by
the quantum mechanical predictions in the case of real experiments.
First we need to prove the following algebraic theorem.

{\it Theorem:} Given ten non-negative real numbers
$x_{1}^{+}$, $x_{1}^{-}$, $x_{2}^{+}$, $x_{2}^{-}$,
$y_{1}^{+}$, $y_{1}^{-}$, $y_{2}^{+}$, $y_{2}^{-}$, $U$ and $V$
such that
$x_{1}^{+}, x_{1}^{-},
x_{2}^{+}, x_{2}^{-} \leq U$,
and
$y_{1}^{+}, y_{1}^{-},
y_{2}^{+}, y_{2}^{-} \leq V$,
then the following inequality always holds:
\begin{eqnarray}{\nonumber}
Z &=& x_{1}^{+}y_{1}^{+}
+x_{1}^{-}y_{1}^{-}
-x_{1}^{+}y_{1}^{-}
-x_{1}^{-}y_{1}^{+}
+y_{2}^{+}x_{1}^{+}
+y_{2}^{-}x_{1}^{-} \\ \nonumber
&-&y_{2}^{+}x_{1}^{-}
-y_{2}^{-}x_{1}^{+}
+y_{1}^{+}x_{2}^{+}
+y_{1}^{-}x_{2}^{-}
-y_{1}^{+}x_{2}^{-}
-y_{1}^{-}x_{2}^{+}
-2x_{2}^{+}y_{2}^{+} \\
&-&2x_{2}^{-}y_{2}^{-}
+Vx_{2}^{+}+Vx_{2}^{-}
+Uy_{2}^{+}+Uy_{2}^{-}+UV
\ge 0.
\end{eqnarray}
{\it Proof}:
Calling $A=y_{1}^{+}-y_{1}^{-}$, we write the function $Z$ as
\begin{eqnarray} {\nonumber}
Z&=&
x_{2}^{+}
\left(-2y_{2}^{+} + A + V \right )
+ x_{2}^{-}
\left(-2y_{2}^{-} - A + V \right ) \\
&+&\left( x_{1}^{+} - x_{1}^{-} \right )
\left(A + y_{2}^{+}-y_{2}^{-} \right )
+ U y_{2}^{+} +  U y_{2}^{-}
+UV.
\end{eqnarray}

\noindent We consider the following eight cases:
\\
(1) First assume

\vspace{0.3 cm}
$\left \{
\begin{array}{c}
-2y_{2}^{+} + A  + V \ge 0,\\
-2y_{2}^{-} - A + V \ge 0,\\
A + y_{2}^{+}-y_{2}^{-}\ge 0.
\end{array} \right.$
\vspace{0.4 cm}

\noindent The function $Z$ is minimized if
$x_{2}^{+}=0, x_{2}^{-}=0$, and
$ x_{1}^{+} - x_{1}^{-} =-U$. Thus
\begin{eqnarray} {\nonumber}
Z &\ge&
-U \left(A +y_{2}^{+} - y_{2}^{-} \right )
+U y_{2}^{+} +  U y_{2}^{-}
+UV \\
&=&U\left(-A + 2y_{2}^{-} + V\right ).
\end{eqnarray}
Since $V \ge A$ and $y_{2}^{-} \ge 0$, $Z \ge 0$.
\vspace{0.7 cm}
\\
(2) Next assume
$\left \{
\begin{array}{c}
-2y_{2}^{+} + A + V < 0,\\
-2y_{2}^{-} - A + V \ge 0,\\
A + y_{2}^{+}-y_{2}^{-}\ge 0.
\end{array} \right.$
\vspace{0.4 cm}

\noindent The function $Z$ is minimized if
$x_{2}^{+}=U, x_{2}^{-}=0$, and
$ x_{1}^{+} - x_{1}^{-} =-U$. Thus
\begin{eqnarray} {\nonumber}
Z &\ge&
U
\left(-2y_{2}^{+} + A + V \right ) -
U \left(A + y_{2}^{+}-y_{2}^{-} \right )
+ U y_{2}^{+} +  U y_{2}^{-}
+UV \\
&=&2U\left(V+y_{2}^{-}-y_{2}^{+}\right ).
\end{eqnarray}
Since $V \ge y_{2}^{+}$, and $y_{2}^{-} \ge 0$, $Z \ge 0$.
\vspace{0.7 cm}
\\
(3) Next assume
$\left \{
\begin{array}{c}
-2y_{2}^{+} + A + V \ge 0,\\
-2y_{2}^{-} - A + V < 0,\\
A + y_{2}^{+}-y_{2}^{-}\ge 0.
\end{array} \right.$
\vspace{0.4 cm}

\noindent The function $Z$ is minimized if
$x_{2}^{+}=0, x_{2}^{-}=U$, and
$ x_{1}^{+} - x_{1}^{-} =-U$. Thus
\begin{eqnarray} {\nonumber}
Z &\ge&
U
\left(-2y_{2}^{-} - A + V \right ) -
U \left(A + y_{2}^{+}-y_{2}^{-} \right )
+ U y_{2}^{+} +  U y_{2}^{-}
+UV \\
&=&2U\left (V - A \right).
\end{eqnarray}
Since $V \ge  A$, $Z \ge 0$.
\vspace{0.7 cm}
\\
(4) Next assume
$\left \{
\begin{array}{c}
-2y_{2}^{+} + A + V \ge 0,\\
-2y_{2}^{-} - A + V \ge 0,\\
A + y_{2}^{+}-y_{2}^{-} < 0.
\end{array} \right.$
\vspace{0.4 cm}

\noindent The function $Z$ is minimized if
$x_{2}^{+}=0, x_{2}^{-}=0$, and
$ x_{1}^{+} - x_{1}^{-} =U$. Thus
\begin{eqnarray} {\nonumber}
Z &\ge&
U \left(A + y_{2}^{+}-y_{2}^{-} \right )
+ U y_{2}^{+} +  U y_{2}^{-}
+UV \\
&=&U\left(A + 2y_{2}^{+} + V\right ).
\end{eqnarray}
Since $V \ge A$ and $y_{2}^{+} \ge 0$, $Z \ge 0$.
\vspace{0.7 cm}
\\
(5) Next assume
$\left \{
\begin{array}{c}
-2y_{2}^{+} + A + V < 0,\\
-2y_{2}^{-} - A + V < 0,\\
A + y_{2}^{+}-y_{2}^{-}\ge 0.
\end{array} \right.$
\vspace{0.4 cm}

\noindent The function $Z$ is minimized if
$x_{2}^{+}=U, x_{2}^{-}=U$, and
$ x_{1}^{+} - x_{1}^{-} =-U$. Thus
\begin{eqnarray} {\nonumber}
Z &\ge&
U
\left(-2y_{2}^{+} + A + V \right )
+U
\left(-2y_{2}^{-} - A + V \right )
-U \left(A + y_{2}^{+}-y_{2}^{-} \right )\\ \nonumber
&+& U y_{2}^{+} +  U y_{2}^{-}
+UV \\ 
&=&U\left(-2y_{2}^{+} -A +3V \right ).
\end{eqnarray}
Since $V \ge A$ and $V \ge y_{2}^{+}$, $Z \ge 0$.
\vspace{0.7 cm}
\\
(6) Next assume
$\left \{
\begin{array}{c}
-2y_{2}^{+} + A + V < 0,\\
-2y_{2}^{-} - A + V \ge 0,\\
A + y_{2}^{+}-y_{2}^{-} < 0.
\end{array} \right.$
\vspace{0.4 cm}

\noindent The function $Z$ is minimized if
$x_{2}^{+}=U, x_{2}^{-}=0$, and
$ x_{1}^{+} - x_{1}^{-} =U$. Thus
\begin{eqnarray} {\nonumber}
Z &\ge&
U\left(-2y_{2}^{+} + A + V \right )+
U \left(A + y_{2}^{+}-y_{2}^{-} \right )
+ U y_{2}^{+} +  U y_{2}^{-}
+UV \\ 
&=&2U\left(A + V \right ).
\end{eqnarray}
Since $V \ge A$, $Z \ge 0$.
\vspace{0.7 cm}
\\
(7) Next assume
$\left \{
\begin{array}{c}
-2y_{2}^{+} + A + V \ge 0,\\
-2y_{2}^{-} - A + V < 0,\\
A + y_{2}^{+}-y_{2}^{-} < 0.
\end{array} \right.$
\vspace{0.4 cm}

\noindent The function $Z$ is minimized if
$x_{2}^{+}=0, x_{2}^{-}=U$, and
$ x_{1}^{+} - x_{1}^{-} = U$. Thus
\begin{eqnarray} {\nonumber}
Z &\ge&
U
\left(-2y_{2}^{-} - A + V \right )
+U \left(A + y_{2}^{+}-y_{2}^{-} \right )
+ U y_{2}^{+} +  U y_{2}^{-}
+UV \\ 
&=&2U\left( y_{2}^{+}-y_{2}^{-} + V \right ).
\end{eqnarray}
Since $V \ge y_{2}^{-}$ and $y_{2}^{+} \ge 0$, $Z \ge 0$.
\vspace{0.7 cm}
\\
(8) Finally assume
$\left \{
\begin{array}{c}
-2y_{2}^{+} + A + V < 0,\\
-2y_{2}^{-} - A + V < 0,\\
A + y_{2}^{+}-y_{2}^{-} < 0.
\end{array} \right.$
\vspace{0.3 cm}

\noindent The function $Z$ is minimized if
$x_{2}^{+}=U, x_{2}^{-}=U$, and
$ x_{1}^{+} - x_{1}^{-} =U$. Thus
\begin{eqnarray} {\nonumber}
Z &\ge&
U
\left(-2y_{2}^{+} + A +V \right )
+U
\left(-2y_{2}^{-} - A + V \right )
+U
\left(A + y_{2}^{+}-y_{2}^{-} \right ) \\ \nonumber
&+& U y_{2}^{+} +  U y_{2}^{-}
+UV \\
&=& U\left(-2y_{2}^{-} + A + 3V \right ).
\end{eqnarray}
Since $V \ge A$ and $V \ge y_{2}^{-}$, $Z \ge 0$,
and the theorem is proved.

Now let $\mbox {\boldmath $a$ ($b$)}$
and $\mbox {\boldmath $a'$ ($b'$)}$
be two arbitrary orientation of the first
(second) polarizer, and let
\begin{eqnarray}{\nonumber}
x_{1}^{\pm}&=&p^{\;\pm}(\mbox{\boldmath $a$} \mid \lambda), \qquad
x_{2}^{\pm}=p^{\;\pm}(\mbox{\boldmath $a'$}|\lambda), \\ 
y_{1}^{\pm}&=&p^{\;\pm}(\mbox{\boldmath $b$}|\lambda), \qquad
y_{2}^{\pm}=p^{\;\pm}(\mbox{\boldmath $b'$}|\lambda).
\end{eqnarray}

\noindent Obviously for each value of $\lambda$, we have
\begin{eqnarray}{\nonumber}
p^{\;\pm}(\mbox{\boldmath $a$} \mid \lambda) \leq 1, \qquad
p^{\;\pm}(\mbox{\boldmath $a'$} \mid \lambda) \leq 1,\\ 
p^{\;\pm}(\mbox{\boldmath $b$} \mid \lambda) \leq 1, \qquad
p^{\;\pm}(\mbox{\boldmath $b'$} \mid \lambda) \leq 1.
\end{eqnarray}

\noindent Inequalities ($4$) and ($15$) yield
\begin{eqnarray} {\nonumber}
&&p^{+}(\mbox{\boldmath $a$} \mid \lambda) \,
p^{+}(\mbox{\boldmath $b$} \mid \lambda)
+p^{-}(\mbox{\boldmath $a$} \mid \lambda) \,
p^{-}(\mbox{\boldmath $b$} \mid \lambda)
-p^{+}(\mbox{\boldmath $a$} \mid \lambda) \,
p^{-}(\mbox{\boldmath $b$} \mid \lambda)  \\ \nonumber
&&- \,p^{-}(\mbox{\boldmath $a$} \mid \lambda) \, 
p^{+}(\mbox{\boldmath $b$} \mid \lambda)
+p^{+}(\mbox{\boldmath $b'$} \mid \lambda) \,
p^{+}(\mbox{\boldmath $a$} \mid \lambda) 
+p^{-}(\mbox{\boldmath $b'$} \mid \lambda) \,
p^{-}(\mbox{\boldmath $a$} \mid \lambda) \\ \nonumber
&&- \, p^{+}(\mbox{\boldmath $b'$} \mid \lambda) \,
p^{-}(\mbox{\boldmath $a$} \mid \lambda) -
p^{-}(\mbox{\boldmath $b'$} \mid \lambda) \,
p^{+}(\mbox{\boldmath $a$} \mid \lambda) +
p^{+}(\mbox{\boldmath $b$} \mid \lambda) \,
p^{+}(\mbox{\boldmath $a'$} \mid \lambda) \\ \nonumber
&&+ \, p^{-}(\mbox{\boldmath $b$} \mid \lambda) \,
p^{-}(\mbox{\boldmath $a'$} \mid \lambda) -
p^{+}(\mbox{\boldmath $b$} \mid \lambda) \,
p^{-}(\mbox{\boldmath $a'$} \mid \lambda) 
-p^{-}(\mbox{\boldmath $b$} \mid \lambda) \,
p^{+}(\mbox{\boldmath $a'$} \mid \lambda) \\ \nonumber
&&- \, 2p^{+}(\mbox{\boldmath $a'$} \mid \lambda) \, 
p^{+}(\mbox{\boldmath $b'$} \mid \lambda)-
2p^{-}(\mbox{\boldmath $a'$} \mid \lambda) \,
p^{-}(\mbox{\boldmath $b'$} \mid \lambda)+ 
p^{+}(\mbox{\boldmath $a'$} \mid \lambda) \\
&&+ \, p^{-}(\mbox{\boldmath $a'$} \mid \lambda)
+p^{+}(\mbox{\boldmath $b'$} \mid \lambda) \,
+p^{-}(\mbox{\boldmath $b'$} \mid \lambda) \ge -1.
\end{eqnarray}

\noindent Multiplying both sides of $(16)$
by $p \, (\lambda)$, integrating over $\lambda$ and
using Eqs. $(2)$, we obtain
\begin{eqnarray} {\nonumber}
&&p^{+ +}(\mbox{\boldmath $a,\, b$}) 
+p^{- \,-}(\mbox{\boldmath $a,\, b$}) 
-p^{+ \,-}(\mbox{\boldmath $a,\, b$}) 
-p^{- \,+}(\mbox{\boldmath $a,\, b$})  
+p^{+ +}(\mbox{\boldmath $b',\, a$})+ \\ \nonumber
&&p^{- \,-}(\mbox{\boldmath $b',\, a$})
-p^{+ \,-}(\mbox{\boldmath $b',\, a$})
-p^{- \,+}(\mbox{\boldmath $b',\, a$}) 
+p^{+ +}(\mbox{\boldmath $b,\, a'$})+ \\ \nonumber
&&p^{- \,-}(\mbox{\boldmath $b,\, a'$}) 
-p^{+ \,-}(\mbox{\boldmath $b,\, a'$}) 
-p^{- \,+}(\mbox{\boldmath $b,\, a'$}) 
-2p^{+ +}(\mbox{\boldmath $a',\, b'$}) - \\
&&2p^{- \,-}(\mbox{\boldmath $a',\, b'$}) 
+p^{+}(\mbox{\boldmath $a'$}) 
+p^{-}(\mbox{\boldmath $a'$}) 
+p^{+}(\mbox{\boldmath $b'$}) 
+p^{-}(\mbox{\boldmath $b'$})
\geq -1.
\end{eqnarray}
All local realistic theories must satisfy inequality $(17)$.

In the atomic cascade experiments, an atom emits two photons in 
a cascade from state $J=1$ to $J=0$. Since the pair of photons
have zero angular momentum, they propagate in the form of spherical
wave. Thus the probability $p \left(\mbox{\boldmath $d_1$},
\mbox{\boldmath $d_2$} \right)$ 
of both photons being simultaneously detected
by two detectors in the directions $\mbox{\boldmath $d_1$}$ and
$\mbox{\boldmath $d_2$}$ is  \cite{4},\cite{6}
\begin{eqnarray}
p \left(\mbox{\boldmath $d_1,\,d_2$} \right)=
\eta^2 \left ({\frac{\displaystyle \Omega}
{\displaystyle 4\pi}}\right) ^2
g \left (\theta,\phi \right ),
\end{eqnarray}
where $\eta$ is the quantum efficiency of the detectors, 
$\Omega$ is the solid angle of the detector, 
$\cos \theta=\mbox{\boldmath $d_1. d_1$}$,
and angle $\phi$ is related to $\Omega$ by
\begin{eqnarray}
\Omega=2 \pi \left (1-\cos \phi \right).
\end{eqnarray}
Finally the function 
$g \left (\theta,\phi \right )$ is the angular correlation function
and in the special case is given by \cite{4}
\begin{eqnarray} 
g \left (\pi, \phi \right ) = 1+
\frac{1}{8} \cos^2 \phi \left (1 + \cos \phi \right)^2.
\end{eqnarray}
If we insert polarizers in front of the detectors, then the
quantum mechanical predictions for
joint detection probabilities are \cite{4},
\cite{6}
\begin{eqnarray} {\nonumber}
p^{+} \left ( \mbox{\boldmath $a$} \right )=
p^{-} \left ( \mbox{\boldmath $a$} \right )=
\eta \left ({\frac{\displaystyle \Omega}{\displaystyle 8 \pi}}
\right), \qquad
p^{+} \left ( \mbox{\boldmath $b$} \right )=
p^{-} \left ( \mbox{\boldmath $b$} \right )=
\eta \left ({\frac{\displaystyle \Omega}{\displaystyle 8 \pi}}
\right), \\ \nonumber
p^{+ \, +} \left ( \mbox{\boldmath $a,\, b$} \right )=
p^{- \, -} \left ( \mbox{\boldmath $a,\, b$} \right )=
\eta^2 \left ({\frac{\displaystyle \Omega}{\displaystyle 8 \pi}}
\right)^2
g \left (\theta,\phi \right )
\left[1+  
\cos 2 \left ( \mbox{\boldmath $a- b$} \right ) \right ], \\  \nonumber
p^{+ \, -} \left ( \mbox{\boldmath $a,\, b$} \right )=
p^{- \, +} \left ( \mbox{\boldmath $a,\, b$} \right )=
\eta^2 \left ({\frac{\displaystyle \Omega}
{\displaystyle 8 \pi}}\right)^2
g \left (\theta,\phi \right )
\left[1- 
\cos 2\left ( \mbox{\boldmath $a- b$} \right ) \right].
\\
\end {eqnarray}

In experiments which are feasible with present technology [5,12],
because $\Omega \ll 4 \pi$,
only a very small fraction of photons are detected 
Thus inequality
$(17)$ can not be used to test the violation of Bell's 
inequality. We now
state a supplementary assumption, and we
show that this assumption is sufficient to make these
experiments (where $\Omega \ll 4 \pi$)
applicable as a test of local theories
(it is important to emphasize that
a supplementary assumption is required primarily
because the solid angle covered by the
aperture of the apparatus,
$\Omega$, is  much less than $4 \pi$ and not because the
efficiency of the detectors, $\eta$, is much smaller than $1$. In
fact in the previous experiments (Ref. $12$),
the efficiency of detectors were larger than
$90 \%$.
However, because
$\Omega \ll  4 \pi$, all previous experiments needed
supplementary assumptions to test locality).
The supplementary assumption is:
For every emission $\lambda$, the detection probability 
by detector $D^{+}$ (or $D^-$) 
is {\it less than or equal} to the sum  of detection probabilities
by detectors $D^{+}$ and $D^-$ when the polarizer
is set along any {\it arbitrary} axis.
If we let $\mbox{\boldmath $r$}$ be an an {\it arbitrary} direction
of the 
first or second polarizer,
then the above supplementary assumption
may be translated into the 
following inequalities
\begin{eqnarray} {\nonumber}
p^{\;+}(\mbox{\boldmath $a$} \mid \lambda) \leq 
p^{\;+}(\mbox{\boldmath $r$} \mid \lambda)+
p^{\;-}(\mbox{\boldmath $r$} \mid \lambda), \qquad
p^{\;-}(\mbox{\boldmath $a$} \mid \lambda) \leq 
p^{\;+}(\mbox{\boldmath $r$} \mid \lambda)+
p^{\;-}(\mbox{\boldmath $r$} \mid \lambda), \\ \nonumber
p^{\;+}(\mbox{\boldmath $a'$} \mid \lambda) \leq 
p^{\;+}(\mbox{\boldmath $r$} \mid \lambda)+
p^{\;-}(\mbox{\boldmath $r$} \mid \lambda), \qquad
p^{\;-}(\mbox{\boldmath $a'$} \mid \lambda) \leq 
p^{\;+}(\mbox{\boldmath $r$} \mid \lambda)+
p^{\;-}(\mbox{\boldmath $r$} \mid \lambda), \\ \nonumber
p^{\;+}(\mbox{\boldmath $b$} \mid \lambda) \leq 
p^{\;+}(\mbox{\boldmath $r$} \mid \lambda)+
p^{\;-}(\mbox{\boldmath $r$} \mid \lambda), \qquad
p^{\;-}(\mbox{\boldmath $b$} \mid \lambda) \leq 
p^{\;+}(\mbox{\boldmath $r$} \mid \lambda)+
p^{\;-}(\mbox{\boldmath $r$} \mid \lambda), \\  \nonumber
p^{\;+}(\mbox{\boldmath $b'$} \mid \lambda) \leq 
p^{\;+}(\mbox{\boldmath $r$} \mid \lambda)+
p^{\;-}(\mbox{\boldmath $r$} \mid \lambda), \qquad
p^{\;-}(\mbox{\boldmath $b'$} \mid \lambda) \leq 
p^{\;+}(\mbox{\boldmath $r$} \mid \lambda)+
p^{\;-}(\mbox{\boldmath $r$} \mid \lambda).
\\
\end{eqnarray}
Now using relations $(4)$, $(14)$ and $(22)$, and applying
the same argument that
led to inequality $(17)$, we obtain the following inequality
\begin{eqnarray} {\nonumber}
&&\bigg[p^{+ \,+}(\mbox{\boldmath $a,\, b$})
+p^{- \,-}(\mbox{\boldmath $a,\, b$})
-p^{+ \,-}(\mbox{\boldmath $a,\, b$})
-p^{- \,+}(\mbox{\boldmath $a,\, b$}) 
+p^{+ +}(\mbox{\boldmath $b',\, a$})   
+p^{- \,-}(\mbox{\boldmath $b',\, a$})   \\  \nonumber
&&-p^{+ \,-}(\mbox{\boldmath $b',\, a$}) 
-p^{- \,+}(\mbox{\boldmath $b',\, a$}) 
+p^{+ \,+}(\mbox{\boldmath $b,\, a'$}) 
+p^{- \,-}(\mbox{\boldmath $b,\, a'$})
-p^{+ \,-}(\mbox{\boldmath $b,\, a'$}) \\  \nonumber
&&-p^{- \,+}(\mbox{\boldmath $b,\, a'$}) 
-2p^{+ +}(\mbox{\boldmath $a',\, b'$})
-2p^{- \,-}(\mbox{\boldmath $a',\, b'$})
+p^{+\,+}(\mbox{\boldmath $a',r$})
+p^{+\,-}(\mbox{\boldmath $a',r$}) \\  \nonumber
&&+p^{-\,+}(\mbox{\boldmath $a',r$}) 
+p^{-\,-}(\mbox{\boldmath $a',r$}) 
+p^{+\,+}(\mbox{\boldmath $r,b'$})
+p^{+\,-}(\mbox{\boldmath $r,b'$})
+p^{-\, +}(\mbox{\boldmath $r,b'$})\\  
&&+p^{-\, -}(\mbox{\boldmath $r,b'$}) \bigg ] \, \bigg / \,
\left[p^{+\,+}(\mbox{\boldmath $r,r$}) 
+p^{+\,-}(\mbox{\boldmath $r,r$})
+p^{-\,+}(\mbox{\boldmath $r,r$})
+p^{-\,-}(\mbox{\boldmath $r,r$}) \right ] \geq -1.
\end{eqnarray}
\noindent Note that in the above inequality the
the number of emissions $N$ from the source
(something which can not be measured experimentally, see Eq. (1))
is eliminated from the ratio.
Inequality $(23)$ contains only double-detection
probabilities. Quantum mechanics violates
this inequality
in case of real experiments where the solid angle covered 
by the aperture of the apparatus, $\Omega$, is  much less than
$4 \pi$.

Inequality (23) may be considerably simplified if we invoke some
of the symmetries that are exhibited in atomic-cascade photon
experiments. For a pair of particles in a singlet state, the
quantum mechanical detection probabilities $p^{\pm\, \pm}_{QM}$ and
expected value $E_{QM}$ exhibit the following
symmetry
\begin{eqnarray} \nonumber
p^{\pm\, \pm}_{QM} \,\left (\mbox{\boldmath $a,b$} \right)=
p^{\pm\, \pm}_{QM}
\,\left ( \mid\mbox{\boldmath $a-b$} \mid\right), \qquad
E_{QM} \,\left (\mbox{\boldmath $a,b$} \right)=
E_{QM} \,\left ( \mid\mbox{\boldmath $a-b$} \mid\right).\\
\end{eqnarray}
We assume that the 
local theories also exhibit the same symmetry
\begin{eqnarray}
p^{\pm\, \pm} \,\left (\mbox{\boldmath $a,b$} \right)=
p^{\pm\, \pm} \,\left ( \mid\mbox{\boldmath $a-b$} \mid\right), \qquad
E \,\left (\mbox{\boldmath $a,b$} \right)=
E \,\left ( \mid\mbox{\boldmath $a-b$} \mid\right), 
\end{eqnarray}
where $E \,\left ( \mid\mbox{\boldmath $a-b$} \mid\right)$ is
the expected value of detection probabilities in local realistic
theories and is defined as
\begin{eqnarray} \nonumber
E \,\left (\mid\mbox{\boldmath $a-b$} \mid \right) &= &
p^{+\, +} \,\left (\mid\mbox{\boldmath $a-b$} \mid\right)-
p^{+\, -} \,\left ( \mid\mbox{\boldmath $a-b$} \mid\right) \\
&-&p^{-\, +} \,\left ( \mid\mbox{\boldmath $a-b$} \mid\right)+
p^{-\, -} \,\left ( \mid\mbox{\boldmath $a-b$} \mid\right).
\end{eqnarray}
Note that there is no harm in assuming Eqs.
$(25)$ since they are subject to experimental test (CHSH
\cite {4}, FC \cite{5}, and CH \cite {6}
made the same assumptions).
Using the above symmetry, inequality $(23)$ is simplified to
\begin{eqnarray} \nonumber
&&\bigg [E \,\left (\mid \mbox{\boldmath $a-b$} \mid \right)+
E \,\left (\mid \mbox{\boldmath $b-a'$} \mid \right)+
E \,\left (\mid \mbox{\boldmath $b'-a$} \mid \right)-
2p^{+\, +} \,\left (\mid\mbox{\boldmath $a'-b'$} \mid\right)
-2p^{-\, -} \,\left (\mid\mbox{\boldmath $a'-b'$} \mid\right) \\ 
\nonumber
&&+p^{+\, +} \,\left (\mid\mbox{\boldmath $a'-r$} \mid\right)+
p^{+\, -} \,\left (\mid\mbox{\boldmath $a'-r$} \mid\right)+
p^{-\, +} \,\left (\mid\mbox{\boldmath $a'-r$} \mid\right) 
+p^{-\, -} \,\left (\mid\mbox{\boldmath $a'-r$} \mid\right)\\
\nonumber
&&+p^{+\, +} \,\left (\mid\mbox{\boldmath $r-b'$} \mid\right)+
p^{+\, -} \,\left (\mid\mbox{\boldmath $r-b'$} \mid\right)+
p^{-\, +} \,\left (\mid\mbox{\boldmath $r-b'$} \mid\right)
+p^{-\, -} \,\left (\mid\mbox{\boldmath $r-b'$} \mid\right)
\bigg ] \, \bigg / \, \\ 
&&\bigg [p^{+\,+}\, (0^\circ) +
p^{+\,-}\, (0^\circ) +
p^{-\,+}\, (0^\circ) +
p^{-\,-}\, (0^\circ) 
\bigg ] \geq -1.
\end{eqnarray}
We now take $\mbox{\boldmath $a'$}$ and 
$\mbox{\boldmath $b'$}$ to be along direction
$\mbox{\boldmath $r$}$,
and we take
$\mbox{\boldmath $a$}$, 
$\mbox{\boldmath $b$}$, and $\mbox{\boldmath $a'$}$ to be
three coplanar axes, each making $120^\circ$ with the other
two,
that is we choose the 
the following orientations,
$\mid \mbox{\boldmath $a - b$} \mid=
 \mid \mbox{\boldmath $b' - a$} \mid=
 \mid \mbox{\boldmath $b - a'$} \mid= 120^\circ $
and 
$ \mid \mbox {\boldmath $a'- b'$} \mid= 
\mid \mbox {\boldmath $a'- r$} \mid= 
\mid \mbox {\boldmath $r- b'$} \mid= 0^\circ $.
Furthermore if we define $K$ as
\begin{eqnarray}
K=p^{+\,+}(0^\circ)
+p^{+\,-}(0^\circ)
+p^{-\,+}(0^\circ)
+p^{-\,-}(0^\circ)
\end{eqnarray}
then the above inequality is simplified to
\begin{eqnarray}
\frac {3E \left( 120^\circ \right)
-2p^{+\, -}\left( 0^\circ \right)
-2p^{- \,+}\left( 0^\circ \right)}{K} \geq -1.
\end{eqnarray}
Using the quantum mechanical probabilities
[i.e., Eqs. $(21)$]
inequality $(29)$ becomes
$-1.5 \geq -1$,
which is certainly impossible.
Quantum mechanics therefore violates inequality  $(29)$
by a factor of 1.5, whereas it violates 
CH (or CHSH) inequality by
by a factor of $\sqrt 2$.
Thus the magnitude of violation of inequality $(29)$
is
approximately $20.7\%$ larger than the magnitude of violation of
the previous inequalities [4-10].

Moreover, inequality (29) can be used to test locality 
considerably more simply than CH or CHSH inequality.
CH inequality may 
be written as
\begin{eqnarray}
\frac{3 p \left( \phi \right)- 
p \left( 3 \phi \right)- 
p \left( \mbox{\boldmath $a'$}, \infty \right)- 
p \left( \infty , \mbox{\boldmath $b$} \right)}{
p \left( \infty , \infty \right)} \leq 0.
\end{eqnarray}
The above inequality requires the measurements of five
detection probabilities:
\\
(1) The measurement of detection probability
with both polarizers set along the $22.5^\circ$ axis
[that is $p\left(22.5^\circ \right)$].
\\
(2) The measurement of detection probability
with both polarizers set along the $67.5^\circ$ axis
[that is $p\left(67.5^\circ \right)$].
\\
(3) The measurement of detection probability
with the first polarizer set along
$\mbox{\boldmath $a'$}$ axis and the second polarizer being
removed
[that is 
$ p\left(\mbox{\boldmath $a'$}, \infty \right)$].
\\
(4) The measurement of detection probability
with the first polarizer removed and the second polarizer set along
$\mbox{\boldmath $b$}$ axis
[that is 
$ p\left(\infty , \mbox{\boldmath $b$} \right)$].
\\
(5) The measurement of detection probability
with both polarizers removed [that is 
$ p \left( \infty , \infty \right)$].
\\
In contrast, the inequality derived in this paper
[i.e., inequality $(29)$] requires the measurements of only
two detection probabilities:
\\
(1) The measurement of detection probability
with both polarizers set along the $0^\circ$ axis
[that is $p\left(0^\circ \right)$].
\\
(2) The measurement of detection probability
with both polarizers set along the $120^\circ$ axis
[that is $p\left(120^\circ \right)$].

Inequality $(29)$ is also experimentally simpler than
FC inequality \cite{5}
(it should be noted that FC inequality is derived under the
assumptions that
(i) $p\left(  \mbox{\boldmath $a'$}, \infty \right)$ is
independent of $\mbox{\boldmath $a'$}$,
(ii) $p\left(  \mbox{\boldmath $b'$}, \infty \right)$ is
independent of $\mbox{\boldmath $b'$}$.
These assumptions, however, should be tested experimentally).
FC inequality may be written as
\begin{eqnarray}
\frac{p\left(  22.5^\circ \right)- 
p\left(  67.5^\circ \right)}{
p\left(  \infty , \infty \right)} \leq 0.25.
\end{eqnarray}
The above inequality requires the measurement of at least three
detection probabilities:
\\
(1) The measurement of detection probability
with both polarizers set along the $22.5^\circ$ axis
[that is $p\left(22.5^\circ \right )$].
\\
(2) The measurement of detection probability
with both polarizers set along the $67.5^\circ$ axis
[that is $p\left(67.5^\circ \right )$].
\\
(3) The measurement of detection probability
with both polarizers removed [that is 
$ p\left( \infty , \infty \right)$].
\\
In contrast inequality $(29)$ 
requires the measurements of only 
two detection probabilities.

A final comment is in order about 
the Bell inequality [inequality $(29)$] that
was derived in this paper.
The analysis that led to inequality $(29)$
is not limited to atomic-cascade experiments and
can easily be extended to experiments which use
phase-momentum \cite {13}, or use high energy
polarized protons or $\gamma$ photons [14-15] to test Bell's limit.
For example in the experiment by Rarity and Tapster \cite{13},
instead of inequality $(2)$ of their paper, the following
inequality (i.e., inequality $(29)$ 
using their notations) may be used to test locality:
\begin{eqnarray}
\frac {3E \left ( 120^\circ \right)
-2{\overline C}_{a_3 \,b_4}\left( 0^\circ \right)
-2{\overline C}_{a_4 \,b_3}\left( 0^\circ \right)}
{K} \geq -1
\end{eqnarray}
where ${\overline C} _{a_i \,b_j} \left ( \phi_a \, , \phi_b \right)$
$(i=3,4;j=3,4)$ is the counting rate between detectors $D_{ai}$ and
$D_{bj}$ with phase angles being 
set to $\phi_a \, , \phi_b$ (See Fig. 1 of \cite {13}).
The following set of orientations
$\mbox{\boldmath $(\phi_a,\, \phi_b)$}=
\mbox{\boldmath $(\phi_{b'},\, \phi_a)$}=
\mbox{\boldmath $(\phi_b,\, \phi_{a'})$}=120^\circ $, and
$\mbox{\boldmath $(\phi_{a'},\, \phi_{b'})$}=0^\circ $
leads to the largest violation of
inequality $(32)$.
Using the optimum orientation of phase angles, 
the magnitude of violation of inequality $(32)$ is
approximately $20.7 \%$ larger than the magnitude of violation of
inequality $(2)$ of \cite {13}.
This result can be particularly important
for experiments using phase-momentum to test locality.
Similarly, in high-energy experiments,
inequality $(29)$ can lead to a larger magnitude of violation. 
For example, in spin correlation proton-proton scattering experiments
\cite {15},
inequality $(29)$ leads to a magnitude of violation of
approximately $20.7 \%$ larger than the results reported by
Lamehirachti and Mittig \cite {15}.

In summary, we have demonstrated that
the conjunction of Einstein's locality
[Eq. $(3)$] with a supplementary assumption [inequality $(22)$]
leads to validity of inequality $(29)$
that is sometimes grossly violated
by quantum mechanics.
Inequality $(29)$, which may be called 
{\em strong} inequality \cite{16},
defines an experiment which can actually
be performed with present technology and which does not require
the number of emissions $N$. Quantum mechanics violates this 
inequality by a factor of 1.5, whereas it violates the
previous {\em strong} inequalities (for example CHSH inequality
of $1969$ \cite{4}, or
CH inequality of $1974$ \cite{6})
by a factor of $\sqrt 2$.
Thus the magnitude of violation of the inequality derived in this
paper is
approximately $20.7\%$ larger than the magnitude of violation of
CH (or CHSH) inequality.
Moreover, inequality $(29)$
requires the measurements of
only two detection probabilities
(at polarizere angles $0^\circ$ and $120^\circ$),
whereas CH or CHSH inequality 
requires the measurements of five detection probabilities.
This result can be of considerable significance for the experimental
test of locality where the time during which the source emits
particles is usually very limited and it is
highly desirable to perform the least number of measurements.

\pagebreak

\end{document}